\documentclass[prl,twocolumn,aps]{revtex4-1}
\usepackage{epsfig}

\begin{document}

\title{Linear Response Theory for Hard and Soft Glassy Materials}

\author{Eran Bouchbinder$^1$ and J.S. Langer$^2$}
\affiliation{$^1$Chemical Physics Department, Weizmann Institute of Science, Rehovot 76100, Israel\\$^2$Department of Physics, University of California, Santa Barbara, CA  93106-9530}

%\date{\today}

\begin{abstract}
Despite qualitative differences in their underlying physics,  both hard and soft glassy materials exhibit almost identical linear rheological behaviors. We show that these nearly universal properties emerge naturally in a shear-transformation-zone (STZ) theory of amorphous plasticity, extended to include a broad distribution of internal thermal-activation barriers. The principal features of this barrier distribution are predicted by nonequilibrium, effective-temperature thermodynamics. Our theoretical loss modulus $G''\!(\omega)$ has a peak at the $\alpha$ relaxation rate, and a power law decay of the form $\omega^{-\zeta}$ for higher frequencies, in quantitative agreement with experimental data.
\end{abstract}
\maketitle

Qualitatively different kinds of amorphous materials -- e.g. structural, metallic and colloidal glasses -- exhibit remarkable similarities in their linear rheological properties \cite{10CWJCY, GAUTHIER-04} despite their enormous range of internal dynamics and intrinsic time scales.  In particular, their frequency dependent loss moduli $G''\!(\omega)$ all have peaks that rise near a viscous relaxation rate and drop slowly over many decades of higher frequencies. We show here that this near universality emerges naturally in the shear-transformation-zone (STZ) theory of amorphous plasticity \cite{FL-98, BLP-07-II, JSL-08, FL-10}, generalized to include a distribution of internal barriers heights of the kind found experimentally in 1980 by Argon and Kuo \cite{ARGON-KUO-80}.  We also show that the principal features of this barrier-height distribution are predicted by nonequilibrium, effective-temperature thermodynamics \cite{BLII-09}. Our analysis differs from both soft glassy rheology (SGR) \cite{SOLLICH-97} and from  mode-coupling theory \cite{BRADER-CATES-09}. In particular, we start with a kinematic formulation that is fundamentally different from that used in SGR, and we insist that the elementary shear tranformations be driven by ordinary thermal fluctuations.

The STZ theory assumes that the degrees of freedom of a glassy material can be separated into two weakly coupled subsystems -- the slow configurational degrees of freedom, i.e. the inherent structures, and the fast kinetic-vibrational degrees of freedom. While the latter are generally in  equilibrium with the thermal reservoir, the former are characterized by an effective temperature that may depart from the reservoir temperature \cite{BLII-09}. A fundamental premise of STZ theory is that irreversible shear deformations occur only at rare, localized, two-state, flow defects in the configurational subsystem. In flowing states, these STZ's appear and disappear as the driven system makes transitions between its inherent structures. In jammed states, the STZ's are configurationally frozen, but they still are able to make transitions between their internal orientations in response to ordinary thermal fluctuations or applied stresses.

For spatially homogeneous systems, we may start by assuming for simplicity that the STZ's are oriented only in the $\pm$ directions relative to the shear stress $s$. We then consider STZ's characterized by an internal, thermal  activation barrier $\Delta$. Let the number of $\pm$ STZ's with given $\Delta$ be $N_{\pm}(\Delta)$, and let the total number of (coarse-grained) molecular sites be $N$. The master equation for $N_{\pm}(\Delta)$ is \cite{FL-10}
\begin{eqnarray}
\label{Ndot}
\nonumber
\tau_0\,\dot N_{\pm}(\Delta) &=& R(\pm s_C,\Delta)\,N_{\mp}(\Delta) - R(\mp s_C,\Delta)\,N_{\pm}(\Delta)\cr\\&+& \rho(\theta) \,\left[N_{eq}(\Delta)/2- N_{\pm}(\Delta)\right].
\end{eqnarray}
Here, $\tau_0$ is a fundamental time scale, for example, a vibration period for molecular glasses or a Brownian diffusion time for colloidal suspensions. $R(\pm s_C,\Delta))/\tau_0$ is the rate per STZ for thermally activated transitions between $\pm$ orientations; and $s_C$ is the partial stress acting on the configurational subsystem. In contrast to SGR, we assume that $s_C$ is a coarse-grained stress, determined by external forcing, and that the molecular-level stresses are accounted for implicitly by the
two-state dynamics. The total stress, $s\!=\!s_C \!+\! \eta_K \ast \dot\gamma$, includes (when appropriate) a partial stress acting on the kinetic-vibrational subsystem.  For colloidal suspensions, this  additional stress is a kinetic viscosity;  $\dot\gamma$ denotes the total shear rate and $\eta_K\ast$ denotes a convolution over time.

The two terms in square brackets on the right-hand-side of Eq. (\ref{Ndot}) are the rates at which STZ's are created and annihilated by spontaneous thermal fluctuations. Mechanically generated noise, as used in \cite{JSL-08, FL-10}, is second order in the applied stress and therefore can be neglected in this linear theory. We are making a detailed-balance approximation in which $N_{eq}(\Delta)$ is the value approached by $2N_{\pm}(\Delta)$ in steady-state. The sum over all possible $\Delta$'s is determined by a Boltzmann-like factor $\exp(-e_Z/\chi)$ according to $\sum_\Delta N_{eq}(\Delta)\!=\!N \exp(-e_Z/\chi)$. $\chi$ is the effective disorder temperature in energy units, and $e_Z$ is a typical STZ formation energy. $\rho(\theta)$ is the thermal noise strength, and $\theta\!=\! k_B\,T$ is the bath temperature in units of energy.

The contribution to the rate of deformation coming from STZ transitions with barrier-height $\Delta$ is
\begin{eqnarray}
\label{DplastDelta}
\nonumber
&&\tau_0\,D^{pl}(\Delta) = {v_0\over V}\,\Bigl[R(s_C,\Delta)\,N_-(\Delta)-R(-s_C,\Delta)\,N_+(\Delta)\Bigr]\cr\\ &&= \epsilon_0\,\Lambda(\Delta)\,{\cal C}(s_C,\Delta)\Bigl[{\cal T}(s_C,\Delta) - m(\Delta)\Bigr],
\end{eqnarray}
where $V$ is the volume of the system, and $v_0$ is a molecular volume that sets the size of the plastic strain increment induced by an STZ transition. We expect $\epsilon_0 \equiv N\,v_0/V$ to be a number of the order of unity. As usual \cite{FL-10}, we have defined
\begin{eqnarray}
\label{Lambda-m-def}
&&\Lambda(\Delta) =  {N_+(\Delta) + N_-(\Delta)\over N},~~ m(\Delta) = {N_+(\Delta) - N_-(\Delta)\over N_+(\Delta) + N_-(\Delta)},\nonumber\\
&&{\cal C}(s_C,\Delta) \equiv {1\over 2}\,\Bigl[R(s_C,\Delta) + R(-s_C,\Delta)\Bigr],\nonumber\\
&&{\cal T}(s_C,\Delta) \equiv {R(s_C,\Delta) - R(-s_C,\Delta)\over R(s_C,\Delta) + R(-s_C,\Delta)}.
\end{eqnarray}
According to Eq. (\ref{Ndot}), the equation of motion for $m(\Delta)$ is
\begin{eqnarray}
\label{mdot}
\!\!\!\tau_0\dot m(\Delta) \!=\! 2{\cal C}(s_C,\Delta)\Bigl[\!{\cal T}(s_C,\Delta) \!-\! m(\Delta)\Bigr] \!-\! \rho(\theta)m(\Delta).
\end{eqnarray}

In deriving Eq. (\ref{mdot}), we have assumed that $\dot\Lambda(\Delta)\!\!=\!\!0$.  This means that we are focusing on time scales in which true structural aging is negligible, and in which the linear response of the system is time-translationally invariant. However, the onset of structural aging does play an important role in what follows.  The rate at which STZ's are spontaneously created and annihilated by thermal fluctuations is proportional to $(\rho/\tau_0)\, \exp(-e_Z/\chi)$, which has the form of a conventional activation rate with $\chi$ playing the role of the temperature and $\rho(\theta)/\tau_0$ being the attempt frequency. The kinetic prefactor $\rho(\theta)$ is a super-Arrhenius function of the temperature, associated with the fact that the configurational rearrangements needed to form or annihilate STZ-like defects involve many-body fluctuations that become increasingly complex and unlikely as the temperature decreases. Accordingly, $\rho(\theta)$ is approximately equal to unity at temperatures well above the glass temperature, but becomes very small at lower temperatures, and vanishes below a glass transition temperature if such a transition exists \cite{JSL-XCHAINS-06}.

We now introduce $p(\Delta)$, a normalized distribution over barrier heights such that $\Lambda(\Delta)\!=\!\exp\! (-\,e_Z/\chi)p(\Delta)d\Delta$, and use Eq. (\ref{DplastDelta}) to write the total rate of irreversible deformation in the form
\begin{equation}
\label{Dplast}
D^{pl}\!(s_C)=\frac{\epsilon_0 e^{-\!\case{e_Z}{\chi}}\!\!}{\tau_0} \!\!\int \!d\Delta\,p(\Delta)\,{\cal C}(s_C,\Delta)\,\left[{\cal T}(s_C,\Delta)\!-\!m(\Delta)\right].
\end{equation}
To complete the derivation, we must specify the transition rate $R(s_C,\Delta)$, which determines ${\cal C}(s_C,\Delta)$ and ${\cal T}(s_C,\Delta)$ according to Eqs.(\ref{Lambda-m-def}). In a linear theory ${\cal C}(s_C,\Delta)$ must be independent of $s_C$, and ${\cal T}(s_C,\Delta)$ must be linear in it.  The simplest assumptions for our purposes are
\begin{equation}
\label{rate_factor}
{\cal C}(s_C,\Delta) \simeq \rho_0(\theta,\Delta)\,e^{- \Delta/\theta},\quad
{\cal T}(s_C,\Delta) \simeq {v_0\,s_C\over a_0\,\theta}.
\end{equation}
Here, $\rho_0(\theta,\Delta)$ is a many-body activation prefactor, which is analogous to $\rho(\theta)$, but which does not necessarily vanish at a glass transition. $a_0$ is a dimensionless number of the order of unity. We stress that ${\cal C}(s_C,\Delta)$ in Eq. (\ref{rate_factor}) describes only ordinary thermally activated processes. We then write
$\nu(\Delta) \equiv 2\,\rho_0(\theta,\Delta)\,\exp(- \Delta/\theta)$, $\tilde p(\nu) = -\,p(\Delta)\,d\Delta/d\nu$, and, in most cases, work with the kinetic quantity $\nu$ instead of the energy $\Delta$ as the independent variable.

To compute the linear oscillatory response, we assume that the total shear rate $\dot\gamma$ is simply the sum of elastic and plastic parts:
\begin{equation}
\label{gammadot}
\dot\gamma = \dot s_C / \mu + D^{pl}(s_C),\quad s_C = s -\,\eta_K\ast\dot \gamma,
\end{equation}
where $\mu$ is the shear modulus. We then denote Fourier transforms as functions of frequency $\omega$ by $\hat \gamma$ etc., let $\eta_K\ast\dot \gamma \to i\,\omega\,\hat \eta_K(\omega)\,\hat\gamma$, and use the preceding equations to solve for $G(\omega) = \hat s/\hat\gamma$. The result is:
\begin{equation}
\label{G}
G(\omega) = i\,\omega\,\tau_0\,\mu\,\left[{{\cal N}(\omega)\over i\,\omega\,\tau_0 + \bar\Lambda\,J(\omega)}\right],~~\bar\Lambda= {\epsilon_0 v_0 \mu \over 2 a_0\theta}\,e^{-e_Z/\chi},
\end{equation}
where
\begin{eqnarray}
\label{various}
\nonumber
{\cal N}(\omega) &=& 1 + {i\,\omega\over \mu}\,\hat\eta_K(\omega)  + {\hat\eta_K(\omega)\over \mu\,\tau_0}\,\bar\Lambda\,J(\omega) \ ,\cr\\
J(\omega) &=& \int\,d\nu\,\tilde p(\nu)\,\nu\,\left({i\,\omega\,\tau_0 +\rho \over i\,\omega\,\tau_0 +\rho  + \nu}\right).
\end{eqnarray}
The storage and loss moduli are, respectively, $G'\!=\!{\rm Re}[G]$ and $G''\!=\!{\rm Im}[G]$.

In interpreting these results, we note first that $G(\omega)$ in Eq. (\ref{G}) cannot, in any physically meaningful way, be expressed as an average over Maxwell modes as in SGR \cite{SOLLICH-97}. This feature is a result of our kinematic assumption in Eq. (\ref{gammadot}), plus our assumption that the plastic strain rate appearing there is a sum over independent contributions from the two-state STZ's with different $\Delta$'s. Next, we use Eq. (\ref{G}) to compute the Newtonian viscosity associated with configurational deformations:
\begin{equation}
\label{etaN}
\eta_N = \lim_{\omega \to 0}\,{G(\omega)\over i\,\omega} = {\mu\,\tau_0\over  \bar\Lambda\,J(0)} \,,
\end{equation}
and deduce immediately, without yet knowing $\tilde p(\nu)$, that the low-frequency structure of  $G(\omega)$ occurs approximately in the neighborhood of $\omega_{\alpha} \sim \bar\Lambda\,J(0)/ \tau_0 = \mu / \eta_N$.  Thus, $\omega_{\alpha}$ is the viscous relaxation rate. However, the structure of the $\alpha$ peak depends sensitively on $\tilde p(\nu)$, and the approximation made above cannot replace a full evaluation of $G(\omega)$.

The distribution $p(\Delta)$ is a near-equilibrium property of the configurational subsystem; therefore, it must be determined by the effective temperature $\chi$.  Because $\Delta$ is measured {\it downward} from some reference energy, we postulate, at least for some range of values of $\Delta$, that  $p(\Delta)$ has the form
\begin{equation}
\label{pDeltalow}
p(\Delta) \propto e^{+ b\,\Delta/\chi},
\end{equation}
where $b$ is a dimensionless number. In the limit of small $\Delta$ (large $\nu$), where $\rho_0 \!\simeq\! 1$, Eq. (\ref{pDeltalow}) becomes $\tilde p(\nu)\!\simeq\!\tilde A \nu^{-1-\zeta}$, where $\zeta\!=\! b\,\theta/\chi$ and $\tilde A$ is a normalization factor.

Integrability of $p(\Delta)$ requires that this distribution be cut off at some $\Delta \!> \!\Delta^*$, $\nu\! <\! \nu^*$. To estimate $\nu^*$, we argue that STZ transition rates cannot be slower than the rates at which the STZ's themselves are appearing and disappearing. Therefore, we propose the important relation:
\begin{equation}
\label{nu*}
\nu^* = 2\,\rho_0(\theta,\Delta^*)\,e^{-\,\Delta^*/\theta} = 2\,\rho(\theta)\,e^{-\,e_Z/\chi}.
\end{equation}
Note that we do not need to know $\rho_0$ in order to determine $\nu^*$ from the last expression on the right-hand side of Eq. (\ref{nu*}).

Since the cutoff at $\Delta^*$ cannot be infinitely sharp, we assume that the distribution drops off exponentially, $p(\Delta)\!\propto\! \exp[-\,(\Delta-\Delta^*)/\Delta_1 ]$ for $\Delta \!>\! \Delta^*$, where $\Delta_1$ is an energy that may be of the order of $\theta$.  We combine these limiting behaviors to write
\begin{equation}
\label{pnu}
\tilde p(\nu) \simeq {\tilde A\over \nu\,\left[(\nu/\nu^*)^{\zeta} + (\nu^*/\nu)^{\zeta_1}\right]} \ ,
\end{equation}
with $\zeta_1\!=\!\theta/\Delta_1 \! \sim \!1$. This is an oversimplified, three-parameter representation of $\tilde p(\nu)$, with two exponents $\zeta$ and $\zeta_1$ determining the large-$\nu$ and small-$\nu$ limits respectively, and a single crossover value of $\nu^*$. It is remarkably similar to the experimentally deduced distributions shown in \cite{ARGON-KUO-80}.  The parameter $\zeta$ controls the high-frequency behavior of $G(\omega)$. A simple scaling analysis for $\omega\tau_0\!\gg\!\rho$ and $\nu^*\!\ll\!2$ predicts that $G''\!(\omega) \sim (\omega\,\tau_0)^{-\zeta}$ above the $\alpha$ peak.

We now compare our theoretical results to experimental data on the rheology of both hard and soft glasses. We first focus on the oscillatory response of structural and metallic glasses, for which $\tau_0$ is of the order of picoseconds, $\omega\tau_0\!\ll\! 1$, and $\eta_K$ is negligible. The interesting behavior occurs at temperatures near or slightly above the glass temperature.

Our principal sources of information about the oscillatory responses of structural and metallic glasses are the papers by Gauthier et al., in particular \cite{GAUTHIER-04}. These authors show that the functions $G(\omega)$, for a wide variety of noncrystalline materials at their glass temperatures, have very similar behaviors. Specifically, the loss modulus $G''\!(\omega)$ has a broad peak at $\omega_{\alpha}$ and drops off at high frequencies like $\omega^{-\zeta}$ as predicted above. For metallic glasses, Gauthier et al. find $\zeta\!\simeq\! 0.4$.

%%%%%%%%%%%%% FIGURE 1 %%%%%%%%%%%%
\begin{figure}[here]
\centering \epsfig{width=.45\textwidth,file=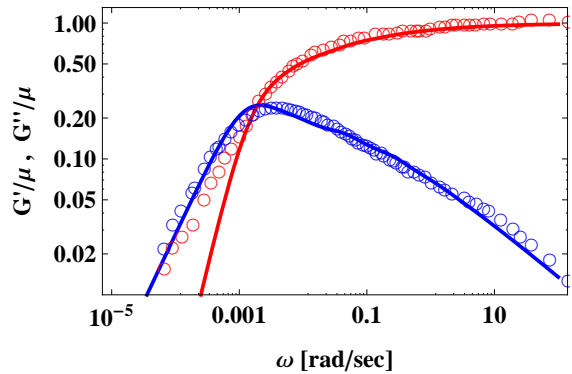} \caption{(Color online) Experimental data for Vitreloy 4 and theoretical comparisons for the storage modulus $G'(\omega)$ (red) and the loss modulus $G''\!(\omega)$ (blue). The data points were extracted from Fig. 2 in \cite{GAUTHIER-04}, where very similar curves for oxide and polymeric glasses can be found.} \label{G-BMG}
\end{figure}
%%%%%%%%%%%%%%%%%%%%%%%%%%%%%%%%%%%%%

In Fig. \ref{G-BMG}, we show  $G'(\omega)/\mu$ and $G''\!(\omega)/\mu$ as predicted by Eq. (\ref{G}), along with data from Fig. 2 of \cite{GAUTHIER-04} for the metallic glass Vitreloy 4 at its glass temperature $T_g$. In estimating the theoretical parameters, we have used $T_g \!\simeq\! 600$K, $\tau_0\!\simeq\! 2 \!\times\! 10^{-12}$sec and $\mu\!\simeq\! 50$ GPa. To make approximations for the other parameters in Eq. (\ref{G}), we note that, if the volume $v_0$ is of the order of a few cubic nanometers, then the ratio $v_0\,\mu /\theta_g$ (the prefactor of $\bar{\Lambda}$ in Eq. (\ref{G})) is approximately $10^4$. Then, to estimate $\Lambda(\chi\!\simeq\!\theta_g)\! \sim\! \exp(-e_Z/\theta_g)$, we assume that $\theta_g$ is the same as the steady-state value of the effective temperature, usually denoted by $\chi_0$, for systems driven persistently at shear rates much slower than $\tau_0^{-1}$.  The ratio $\chi_0/e_Z$ may be a universal quantity; it usually turns out to be in the range $0.1\!-\!0.2$. We therefore estimate $\theta_g/e_Z \!\sim\! 0.15$, implying that $\Lambda(\theta_g)\!\sim\!10^{-3}$, and thus that $\bar\Lambda \!\sim\! 10$. Then Eq. (\ref{nu*}) tells us that $\nu^* \!\simeq\! 10^{-3}\rho(\theta_g)$.

The theoretical curves in Fig. \ref{G-BMG} have been computed using  $\zeta_1\!=\!1$, $\rho(\theta_g)/\tau_0\!=\! 1.25\! \times \!10^{-2}$sec$^{-1}$ and $\bar\Lambda\!=\!25$. In effect, we have set $\epsilon_0 / a_0 \!\sim \!2.5$ in Eq. (\ref{G}), which is well within our theoretical uncertainty. It is important to note that these parameters imply that $\nu^*\!\sim\! 10^{-17}$, which is extremely small compared to its upper limit at $\nu\!=\! 2$. So far as we can tell from numerical exploration, this small value of $\nu^*$ is sharply determined by the experimental data. The major discrepancy between the theoretical curve and the experimental data is that the measured $G'(\omega)$ is approximately linear in $\omega$ at low frequencies, instead of being proportional to $\omega^2$ as predicted by our theory and by Maxwell models.

We turn now to soft glasses, in particular, to thermosensitive colloidal suspensions, whose rheology differs from that of structural and metallic glasses in at least two important respects.  First, in colloidal systems, the approach to jamming near a glass transition is controlled more sensitively by the volume fraction than by the temperature. Second, the microscopically short molecular vibration period in structural glasses is replaced in colloids by the very much longer time scale for Brownian motion of the particles. As a result, the high-frequency cutoff at $\nu\!=\!2$ is probed in rheological experiments, and the kinetic viscosity $\hat\eta_K(\omega)$ is relevant at accessibly high values of $\omega$.

%%%%%%% FIGURE 2 %%%%%%%%%%%%%%%%%%%%
\begin{figure}
\centering \epsfig{width=.45\textwidth,file=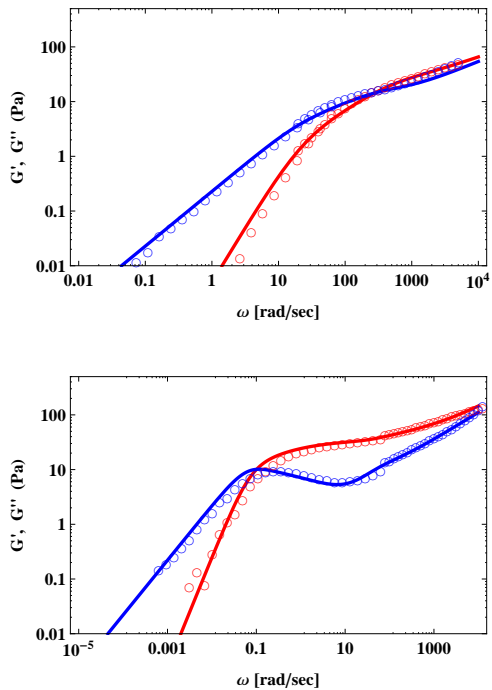}
\caption{Experimental data and theoretical comparisons for the storage modulus $G'(\omega)$ (red) and the loss modulus $G''\!(\omega)$ (blue) for two different suspensions of thermosensitive particles, as reported in \cite{SIEBENBURGER-09}. The values of the parameters for the top and bottom panels, respectively, are: $\rho\!=\!0.04,\, 3 \times 10^{-4}$; $\nu^*\!=\! 0.001,\,10^{-3}\rho$; $\zeta\!=\!1.0,\, 0.5$; $\bar\Lambda\!=\!200,\, 40$; $\mu\!=\!12,\, 35$ Pa; and $\tau_K\!=\! 0.004,\,0.002$ sec. In all cases, $\zeta_1\!=\!1$ and $c\!=\!0.1$.}
\label{G-Colloid}
\end{figure}
%%%%%%%%%%%%%%%%%%%%%%%%%%%%%%%%%%%%%

In order to evaluate $G(\omega)$ for colloidal suspensions, we need an expression for the frequency-dependent kinetic viscosity $\hat\eta_K(\omega)$. Here, we follow \cite{LIONBERGER-RUSSEL-94} and write $\hat\eta_K(\omega)\!=\!\mu\tau_K (c \!+\! i\omega\tau_K)^{-1/2}$, where $\tau_K$ is a viscous time scale, and $c$ is a dimensionless constant that we have added in order to regularize this formula at small $\omega$.

In Fig. \ref{G-Colloid} we show two examples of how the STZ theory developed here is capable of reproducing the experimental results of Siebenburger et al. \cite{SIEBENBURGER-09}. These authors explored a range of effective volume fractions $\phi_{e\!f\!f}$ and a wide range of frequencies $\omega$ by using suspensions of thermosensitive particles (polystyrene cores with attached networks of thermosensitive isopropylacrylamide molecules). The Brownian time scale in these experiments is $\tau_0\!\simeq\!0.003$ sec. The volume fraction for the top panel is $\phi_{e\!f\!f}\!=\! 0.518$, while that for the bottom panel is $0.600$. The theoretical parameters, deduced by fitting the data, are listed in the figure caption.

The trends are interesting. The example in the top panel of Fig. \ref{G-Colloid} is a system whose relatively small volume fraction puts it well away from the glass transition. It is effectively a liquid; $\rho\!=\!0.04$ means that there is relatively little super-Arrhenius suppression of the structural relaxation rate. In contrast, the system in the bottom panel is much more glassy, and $\rho$ decreases significantly. The shear modulus $\mu$ increases and the viscous time scale $\tau_K$ decreases slightly as the system becomes stiffer. $\bar\Lambda$ is very large for the liquidlike example in the top panel, implying that the STZ density is large in this system. On the other hand, the value of $\bar\Lambda$, and the relation between $\nu^*$ and $\rho$ for the glassy system in the bottom panel are comparable to the estimates for the bulk metallic glass in the preceding discussion -- despite the fact that the underlying time scales for these systems differ by nine orders of magnitude.

The linear oscillatory measurements discussed here are sensitive probes of the internal structure of glassy materials, revealing the broad range of activation mechanisms that occur within them and the relation between these mechanisms and the effective-temperature thermodynamics of glassy disorder.  One of the deepest questions that we have not addressed here, however, is the relation between internal STZ dynamics and true structural aging, in which the slow configurational degrees of freedom relax toward thermodynamic equilibrium.  We have begun to address the latter issues and will report on them in a forthcoming publication  \cite{LB-11}.

\begin{acknowledgments}

We thank M. Siebenburger for sending us the data shown in Fig. \ref{G-Colloid}. JSL was supported in part by the Division of Materials Science and Engineering, Office of Basic Energy Sciences, Department of Energy, DE-AC05-00OR-22725, through a subcontract from Oak Ridge National Laboratory. EB was supported in part by the Harold Perlman Family Foundation and in part by a grant from the Robert Rees Applied Research Fund.

\end{acknowledgments}

\end{document}